\documentclass{ifacconf}

\usepackage{cite}
\usepackage{amsmath,amssymb,amsfonts}
\usepackage{algorithmic}
\usepackage{graphicx}
\usepackage{textcomp}
\usepackage{xcolor}
\usepackage{mathrsfs}
\usepackage{psfrag}
\usepackage{cancel}
\usepackage{verbatim}
\usepackage{mathabx}
\usepackage{colonequals}

\usepackage{circuitikz}
\usetikzlibrary{decorations.pathreplacing,calligraphy,patterns}

\makeatletter
\let\old@ssect\@ssect 
\usepackage{natbib}
\usepackage{hyperref}
\hypersetup{colorlinks,linkcolor={blue},citecolor={blue},urlcolor={red}}
\def\@ssect#1#2#3#4#5#6{%
  \NR@gettitle{#6}
  \old@ssect{#1}{#2}{#3}{#4}{#5}{#6}
}
\makeatother

\def\mb{\mathbb}
\def\wc{\widecheck}

\newtheorem{definition}{Definition}
\newtheorem{theorem}{Theorem}

\newtheorem{lemma}{Lemma}
\newtheorem{remark}{Remark}
\newtheorem{corollary}{Corollary}
\newtheorem{assumption}{Assumption}

\begin{document}
\sloppy
\begin{frontmatter}
\title{\LARGE \bf Discrete-time Negative Imaginary Systems from ZOH Sampling\thanksref{footnoteinfo}}
\thanks[footnoteinfo]{This work was supported by the Australian Research Council under grants DP190102158 and DP230102443.}
\author[ANU]{Kanghong Shi}
\author[ANU]{Ian R. Petersen, \textit{Life Fellow, IEEE}}
\author[ANU]{Igor G. Vladimirov}

\address[ANU]{School of Engineering, College of Engineering, Computing and Cybernetics, Australian National University, Canberra, Acton, ACT 2601, Australia.
        {\tt kanghong.shi@anu.edu.au}, {\tt ian.petersen@anu.edu.au}, {\tt igor.vladimirov@anu.edu.au}.}

\maketitle
\thispagestyle{plain}
\pagestyle{plain}

\begin{abstract}
A new definition of discrete-time negative imaginary (NI) systems is provided. This definition characterizes the dissipative property of a zero-order hold sampled continuous-time NI system. Under some assumptions, asymptotic stability can be guaranteed for the closed-loop interconnection of an NI system and an output strictly negative imaginary system, with one of them having a one step advance. In the case of linear systems, we also provide necessary and sufficient frequency-domain and LMI conditions under which the definition is satisfied. Also provided is a simple DC gain condition for the stability results in the linear case.
\end{abstract}

\end{frontmatter}

\section{INTRODUCTION}
Negative imaginary (NI) systems theory was introduced in \cite{lanzon2008stability} and \cite{petersen2010feedback} to address a robust control problem for flexible structures with colocated force actuators and position sensors. As negative velocity feedback control \cite{brogliato2007dissipative} may not be always applicable, NI systems theory provides an alternative approach where positive feedback control is used. Roughly speaking, a transfer function matrix $F(s)$ is said to be NI if it is stable and $j(F(j\omega)-F(j\omega)^*)\geq 0$ for all $\omega \geq 0$. Under some assumptions, the positive feedback interconnection of an NI system with transfer function matrix $F(s)$ and a strictly negative imaginary (SNI) system with transfer function matrix $R(s)$ is internally stable if and only if the DC loop gain has all its eigenvalues less than unity; i.e., $\lambda_{max}(F(0)R(0))<1$. NI systems theory has attracted significant attention since it was introduced in 2008; e.g., see \cite{xiong2010negative,song2012negative,mabrok2014generalizing,wang2015robust,bhowmick2017lti}. It has been applied in many fields such as nano-positioning control \cite{mabrok2013spectral,das2015multivariable,nikooienejad2021convex,shi2023negative} and the control of lightly damped structures \cite{cai2010stability,rahman2015design,bhikkaji2011negative}.

NI systems theory was extended to nonlinear systems in \cite{ghallab2018extending,shi2021robust,shi2023output} as a dissipativity type property. A nonlinear system is said to be NI if it is dissipative with respect to the supply rate $u^T\dot y$, where $u$ and $y$ are the input and output of the system, respectively. Under some assumptions, the closed-loop interconnection of a nonlinear NI system and a nonlinear output strictly negative imaginary (OSNI) system is asymptotically stable \cite{shi2021robust,shi2023output}. NI systems theory can be regarded as complementary to passivity and positive real (PR) systems theory. An advantage of  NI systems theory is that it can deal with systems having relative degree zero, one and two \cite{shi2021necessary,shi2021negative,dannatt2023strictly}, while passivity and PR systems theory can only deal with systems having relative degree zero or one \cite{brogliato2007dissipative}.

Nowadays, digital computers are almost always used in the control of physical systems \cite{aastrom2013computer}. During the digital control of a continuous-time physical system, a computer reads the output signal of the plant at discrete instants of time. Then the computer generates a discrete-time control signal according to a certain control law. The discrete-time control signal is fed into the plant as a constant continuous-time signal for an entire sampling period until the next control signal is generated. This process is termed zero-order hold (ZOH) sampling. 

For the digital control of NI systems, we establish a stability result in discrete-time for ZOH sampled NI systems. We provide a discrete-time NI system definition such that it is automatically satisfied by ZOH sampled continuous-time NI systems. Also, we propose a feedback control approach for discrete-time NI systems.

Note that prior to this work, a notion of discrete-time NI systems was introduced in \cite{ferrante2017discrete,liu2017properties}. In \cite{ferrante2017discrete,liu2017properties}, discrete-time NI systems are defined by analogy to continuous-time NI systems, which was inspired by the discrete-time PR systems theory \cite{hitz1969discrete}. It is shown in \cite{ferrante2017discrete,liu2017properties} that the continuous-time NI system definition and stability results can be mapped to discrete time using a bilinear transform. However, as pointed out in many papers (see e.g. \cite{de2002preserving,premaratne1994discrete,kawano2023krasovskii,oishi2010passivity}), the discrete-time passive \cite{byrnes1994losslessness} and PR systems \cite{hitz1969discrete} definitions lead to several issues as compared to their original continuous-time analogues. A commonly observed issue is that discretizating a continuous-time passive system using a ZOH device does not in general yield a discrete-time passive system \cite{brogliato2007dissipative}. This issue is also present in the discrete-time NI system definition introduced in \cite{ferrante2017discrete,liu2017properties}.

Also, as mentioned above, the results in \cite{ferrante2017discrete,liu2017properties} were obtained from continuous-time results using a bilinear transform, which is a frequency-domain method and hence is only applicable to linear time-invariant systems. Therefore, it is useful to develop an alternative time-domain formulation of discrete-time NI properties for nonlinear systems with a clearer link to work-energy balance relations.

The contribution of this paper is two-fold: (i) it provides a theory for the digital control of NI systems; (ii) it can be regarded as an alternative discrete-time NI system theory since different methods of continuous-time to discrete-time conversion are used here as compared to \cite{ferrante2017discrete,liu2017properties}. To be specific, we provide a general nonlinear definition of discrete-time NI systems. We introduce a class of systems called discrete-time output strictly negative imaginary (OSNI) systems to be used as controllers for discrete-time NI systems. Note that the discrete-time OSNI property is related to, yet different from the continuous-time OSNI property (see e.g. \cite{shi2021robust,bhowmick2017lti} for a discussion of continuous-time OSNI systems). We show that under some assumptions, the interconnection of an NI system and an OSNI system is asymptotically stable provided that one of their outputs takes a one step advance (which is physically realizable). NI systems and OSNI systems with a one step advance are called step advanced negative imaginary (SANI) systems and step advanced output strictly negative imaginary (SAOSNI) systems, respectively. To be specific, we prove the stability of the interconnection of an NI system and an SAOSNI system and similarly, the stability of the interconnection of an SANI system and an OSNI system, given that certain assumptions are satisfied. We also specialize our results to the case of linear systems. We provide linear matrix inequality (LMI) conditions and frequency-domain conditions under which a system is NI, which are different from, yet related to, the conditions provided in \cite{ferrante2017discrete}. For linear systems, the assumptions required in the stability results reduce to a simple DC gain condition.

The rest of the paper is organized as follows: In Section \ref{section:DT-NI systems}, we provide the new definition of discrete-time NI systems. We also show that discretizing a continuous-time NI system using ZOH sampling yields a discrete-time NI system.  Section \ref{sec:stability} provides the stability results, which are the main contributions of this paper. Section \ref{sec:nonlinear DT-NI for linear} provides conditions for a discrete-time linear system to be NI. In Section \ref{sec:nonlinear DT-NI for linear}, we also specialize the stability results given in Section \ref{sec:stability} to linear systems. Section \ref{sec:conclusion} concludes the paper. 

Notation: The notation in this paper is standard. $\mathbb R$ denotes the set of real numbers. $\mb N$ denotes the set of nonnegative integers. $\mathbb R^{m\times n}$ denotes the space of real matrices of dimension $m\times n$. $A^T$ and $A^*$ denote the transpose and the complex conjugate transpose of a matrix $A$, respectively. $A^{-T}$ denotes the transpose of the inverse of $A$; that is, $A^{-T}=(A^{-1})^T=(A^T)^{-1}$. $\lambda_{max}(A)$ denotes the largest eigenvalue of a matrix $A$ with real spectrum. $\|\cdot\|$ denotes the standard Euclidean norm. For a real symmetric or complex Hermitian matrix $P$, $P>0\ (P\geq 0)$ denotes the positive (semi-)definiteness of a matrix $P$ and $P<0\ (P\leq 0)$ denotes the negative (semi-)definiteness of a matrix $P$. A function $V: \mb R^n \to \mb R$ is called positive definite if $V(0)=0$ and $V(x)>0$ for all $x\neq 0$. A function $V: \mb R^n \to \mb R$ is said to be of class $C^1$ if it continuously differentiable.

\section{DISCRETE-TIME NI SYSTEMS}\label{section:DT-NI systems}
In this section, we provide our definition of discrete-time NI systems and show that it will be automatically satisfied by ZOH sampled continuous-time NI systems.
Consider the following nonlinear discrete-time system
\begin{subequations}\label{eq:H1}
\begin{align}
H_1\colon \quad 	x_{k+1} =&\ f(x_k,u_k),\\
	y_k=&\ h(x_k),
\end{align}	
\end{subequations}
where $f\colon\mathbb R^n \times \mathbb R^p\to \mathbb R^n$ and $h\colon\mathbb R^n \to \mathbb R^p$. Here $u_k,y_k \in \mathbb R^p$ and $x_k\in \mathbb R^n$ are the input, output and state of the system, respectively, at time step $k\in \mathbb N$.

\begin{definition}[discrete-time NI systems]\label{def:DT_NNI}
The system (\ref{eq:H1}) is said to be a discrete-time negative imaginary (NI) system if there exists a continuous positive definite storage function $V\colon \mb R^n \to \mb R$ such that for arbitrary states $x_k$ and inputs $u_k$,
\begin{equation}\label{eq:NNI ineq}
V(x_{k+1})-V(x_{k})\leq u_k^T\left(y_{k+1}-y_{k}\right),	
\end{equation}
for all $k$.
\end{definition}
We show in the following that a continuous-time NI system becomes a discrete-time NI system after ZOH sampling. Consider a continuous-time system
\begin{equation}\label{eq:CT nonlinear system}
\begin{aligned}
\dot x(t) =&\ \mathfrak f(x(t),u(t)),\\
y(t) =&\ \mathfrak h(x(t)),
\end{aligned}
\end{equation}
where $u,y\in \mb R^p$ and $x\in \mb R^n$ are its input, output and state, respectively. Here, $f\colon\mb R^n\times \mb R^p\to \mb R^n$ is a Lipschitz continuous function and $h\colon\mb R^n \to \mb R^p$ is a continuously differentiable function. The continuous-time NI property is defined as follows. Note that we use the definition of NI systems that does not allow for free body motion, as given in \cite{shi2021robust}.

\begin{definition}[continuous-time NI systems]\label{def:CT NI}\cite{shi2021robust}
The system (\ref{eq:CT nonlinear system}) is said to be an NI system if there exists a positive definite storage function $V:\mathbb R^n\to \mathbb R$ of class $C^1$ such that for any locally integrable input $u$ and solution $x(t)$ to (\ref{eq:CT nonlinear system}),
\begin{equation}\label{eq:CT NI inequality}
    \dot V(x(t))\leq u(t)^T\dot y(t),
\end{equation}
for all $t\geq 0$.
\end{definition}

\begin{lemma}
	Suppose a continuous-time system of the form (\ref{eq:CT nonlinear system}) is NI according to Definition \ref{def:CT NI}, then ZOH sampling the system gives a discrete-time NI system satisfying Definition \ref{def:DT_NNI}.
\end{lemma}
\begin{pf}
	Since ZOH sampling is used, then the input of the system (\ref{eq:CT nonlinear system}) remains constant over the time interval $[t_k,t_{k+1})$; i.e., $u(t)=u_k$ for $t\in [t_k,t_{k+1})$. Then, integrating the NI inequality (\ref{eq:CT NI inequality}) over the time interval $[t_k,t_{k+1})$ yields
\begin{equation*}
		\int_{t_k}^{t_{k+1}}\dot V(x(\tau))d\tau \leq \int_{t_k}^{t_{k+1}} u_k^T\dot y(\tau) d \tau.
	\end{equation*}
	That is
	\begin{equation*}
		V(x_{k+1})-V(x_k)\leq u_k^T(y_{k+1}-y_k),
	\end{equation*}
which is the discrete-time NI inequality \eqref{eq:NNI ineq} in Definition \ref{def:DT_NNI}, with $x_k=x(t_k)$, $y_k=y(t_k)$ and similarly defined $x_{k+1}$ and $y_{k+1}$.\hfill $\blacksquare$
\end{pf}

Indeed, inspired by the above discrete-time negative imaginary system definition, the definition of continuous-time NI systems can be equivalently given as follows.
\begin{definition}[continuous-time NI systems]\label{def:CT NI equivalent def}
	The system (\ref{eq:CT nonlinear system}) is said to be an NI system if there exists a positive definite storage function $V\colon \mb R^n \to \mb R$ of class $C^1$ such that for any time $T>0$, initial condition $x(0)$ and any constant input vector $u(t)=v$ for $t\in [0,T]$,
	\begin{equation*}
		V(x(T))-V(x(0))\leq v^T (y(T)-y(0)).
	\end{equation*} 
\end{definition}
An advantage of Definition \ref{def:CT NI equivalent def} is that it does not involve any derivative or integral terms (though it is obtained by integration over $[0,T]$), in comparison to Definition \ref{def:CT NI} and the definition of passive systems \cite{khalil2002nonlinear}.

\begin{remark}
	Definition \ref{def:DT_NNI} does not reduce to the definition of discrete-time NI systems in \cite{ferrante2017discrete} in the linear time-invariant case. This is because \cite{ferrante2017discrete} considers a bilinear transform for conversion of a continuous-time system to a discrete-time system, while this paper considers ZOH sampling. Throughout the remainder of the paper, the term ``NI" refers to the discrete-time NI property as defined in Definition \ref{def:DT_NNI}, unless stated otherwise.
\end{remark}

\section{STABILITY FOR INTERCONNECTED NI SYSTEMS}\label{sec:stability}
In this section, we investigate the stability of the feedback interconnection of NI systems. We apply a class of systems called OSNI systems with a step advance as controllers for NI plants. First, we provide the definition of OSNI systems as follows. Consider a system with the model
\begin{subequations}\label{eq:H2}
\begin{align}
H_2\colon \quad	\widehat x_{k+1} =&\ \widehat f(\widehat x_k, \widehat u_k),\label{eq:H2 state}\\
	\widehat y_k=&\ \widehat h(\widehat x_k),\label{eq:H2 output}
\end{align}	
\end{subequations}
where $\widehat f\colon\mathbb R^m \times \mathbb R^p\to \mathbb R^m$ and $\widehat h\colon\mathbb R^m \to \mathbb R^p$. Here $\widehat u_k, \widehat y_k \in \mathbb R^p$ and $\widehat x_k\in \mathbb R^m$ are the input, output and state of the system, respectively, at time step $k\in \mathbb N$.

\begin{definition}[discrete-time OSNI systems]\label{def:DT-NOSNI}
The system (\ref{eq:H2}) is said to be a discrete-time output strictly negative imaginary (OSNI) system if there exists a continuous positive definite storage function $\widehat V\colon \mb R^m \to \mb R$ and a scalar $\epsilon >0$ such that for arbitrary states $\widehat x_k$ and inputs $\widehat u_k$,
\begin{equation*}
\widehat V(\widehat x_{k+1})-\widehat V(\widehat x_{k})\leq \widehat u_k^T\left(\widehat y_{k+1}-\widehat y_{k}\right)-\epsilon\|\widehat y_{k+1}-\widehat y_{k}\|^2,	
\end{equation*}
for all $k$. In this case, we also say that system (\ref{eq:H2}) is OSNI with degree of output strictness $\epsilon$.
\end{definition}
Note that unlike the discrete-time NI system definition, Definition \ref{def:DT-NOSNI} is not satisfied in general for ZOH sampled continuous-time OSNI systems. The purpose of introducing the output strictness is to guarantee asymptotic stability.

We show in the following that the NI system $H_1$ can be asymptotically stabilized by an auxiliary system associated with the OSNI system $H_2$ in positive feedback. The auxiliary system, denoted by $\widetilde H_2$, is described by (\ref{eq:H2 state}) and 
\begin{equation}\label{eq:tilde H2 output}
	\widetilde y_k = \widehat h (\widehat f (\widehat x_k,\widehat u_k))= \widehat y_{k+1}.
\end{equation}
That is, the auxiliary system takes the output of the next step of the system $H_2$ as its output of the current step. Therefore, for an OSNI system (\ref{eq:H2}), we call the auxiliary system given by (\ref{eq:H2 state}), (\ref{eq:tilde H2 output}) a step advanced output strictly negative imaginary (SAOSNI) system. A formal definition of SAOSNI systems is as follows.

Consider the system 
\begin{subequations}\label{eq:tilde H2}
\begin{align}
\widetilde H_2\colon \quad	\widehat x_{k+1} =&\  \widehat f(\widehat x_k, \widehat u_k),\label{eq:tilde H2 state}\\
	\widetilde y_k=&\ \widetilde h(\widehat x_k,\widehat u_k),
\end{align}	
\end{subequations}
where $\widehat f\colon \mathbb R^m \times \mathbb R^p\to \mathbb R^m$ and $\widetilde h\colon\mathbb R^m\times \mb R^p \to \mathbb R^p$. Here $\widehat u_k, \widetilde y_k \in \mathbb R^p$ and $\widehat x_k\in \mathbb R^m$ are the input, output and state of the system, respectively, at time step $k\in \mathbb N$.
\begin{definition}\label{def:DT-SAOSNI}
	The system \eqref{eq:tilde H2} is said to be a discrete-time step advanced output strictly negative imaginary (SAOSNI) system if there exists a continuous function $\widehat h(\widehat x_k)$ such that:
\begin{enumerate}
		\item $\widetilde h(\widehat x_k,\widehat u_k)=\widehat h(\widehat f(\widehat x_k,\widehat u_k))$;
		\item There exists a continuous positive definite storage function $\widehat V\colon \mb R^m \to \mb R$ and a scalar $\epsilon>0$ such that for arbitrary states $\widehat x_k$ and inputs $\widehat u_k$,
\begin{align*}
\widehat V(\widehat x_{k+1})-\widehat V(\widehat x_{k})\leq &\ \widehat u_k^T\left(\widehat h(\widehat x_{k+1})-\widehat h (\widehat x_{k})\right)\notag\\
&-\epsilon\|\widehat h(\widehat x_{k+1})-\widehat h (\widehat x_{k})\|^2,	
\end{align*}
for all $k$.
\end{enumerate}
\end{definition}

\begin{figure}[h!]
\centering
\psfrag{in_1}{$u_k$}
\psfrag{y_1}{$y_k$}
\psfrag{e}{$\widehat u_k$}
\psfrag{x_h}{$\widetilde y_k$}
\psfrag{plant}{\hspace{0cm}$H_1$}
\psfrag{HIGS}{\hspace{0.2cm}$\widetilde H_2$}
\includegraphics[width=8.5cm]{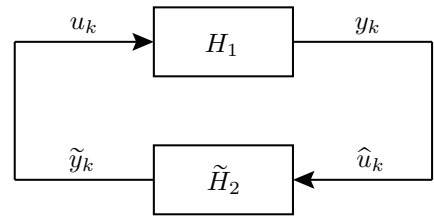}
\caption{Closed-loop interconnection of the NI plant $H_1$ and the SAOSNI controller $\widetilde H_2$.}
\label{fig:interconnection}
\end{figure}

We apply an SAOSNI controller in positive feedback to control an NI plant. The settings of the feedback interconnection is shown in Fig.~\ref{fig:interconnection} and also described by the following equations:
\begin{align*}
\widehat u_k =&\  y_k;\\
u_k =&\ \widetilde y_k.
\end{align*}

\begin{figure}[h!]
\centering
\psfrag{in_1}{$u_k$}
\psfrag{out_1}{$y_k$}
\psfrag{in_2}{\hspace{0.1cm}$\widehat u_k$}
\psfrag{out_2}{\hspace{0.08cm}$\widetilde y_k$}
\psfrag{sys1}{\hspace{0cm}$H_1$}
\psfrag{sys2}{\hspace{0.05cm}$\widetilde H_2$}
\includegraphics[width=8.5cm]{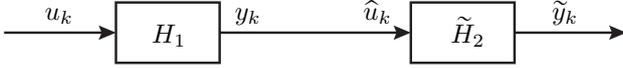}
\caption{Open-loop interconnection of the NI plant $H_1$ and the SAOSNI controller $\widetilde H_2$.}
\label{fig:open-loop1}
\end{figure}

In order to present the stability result, we need to make the following assumptions for the open-loop interconnection of the systems $H_1$ and $\widetilde H_2$ as shown in Fig.~\ref{fig:open-loop1}.

\begin{assumption}\label{assumption1}
	Giving an input $u_k=\overline u$ for the system $H_1$ which is constant for all time steps $k\geq k_c$, we obtain a corresponding output $y_k$. Also, setting the input of the system $\widetilde H_2$ to be $\widehat u_k=y_k$ in the open-loop interconnection shown in Fig.~\ref{fig:open-loop1} with corresponding output $\widetilde y_k$. If $\widetilde y_k=\overline {\widetilde y}$ remains constant for all $k\geq k_c$, and $\overline {\widetilde y}=\overline u$, then
$x_k,\widehat x_k=0$
for all $k\geq k^*$, where $k^*$ is some integer such that $k^*\geq k_c$.
\end{assumption}

We have the following stability result.

\begin{theorem}\label{theorem:stability NI and SAOSNI}
	Consider the NI system $H_1$ given by (\ref{eq:H1}) and the SAOSNI system $\widetilde H_2$ given by (\ref{eq:tilde H2}). Also, suppose Assumption \ref{assumption1} is satisfied and the function
\begin{equation}\label{eq:W NI SAOSNI}
W(x_k,\widehat x_k) = V(x_k)+\widehat V(\widehat x_k)-h(x_k)^T\widehat h(\widehat x_k)
\end{equation}
is continuous and positive definite, where $V(x_k)$ and $\widehat V(\widehat x_k)$ are the storage functions of the systems $H_1$ and $\widetilde H_2$, respectively. Then the closed-loop interconnection of the system $H_1$ and the system $\widetilde H_2$ shown in Fig.~\ref{fig:interconnection} is asymptotically stable.
\end{theorem}
\begin{pf}
Using Lyapunov's direct method \cite{Kalman1960}, consider the increment of the function $W(x_k,\widehat x_k)$ in (\ref{eq:W NI SAOSNI}):
\begin{align*}
	&W(x_{k+1},\widehat x_{k+1})-W(x_{k},\widehat x_{k})\notag\\
	 =&\ V(x_{k+1})+\widehat V(\widehat x_{k+1})-V(x_{k})-\widehat V(\widehat x_{k})-y_{k+1}^T\widehat y_{k+1}\notag\\
	 &+y_{k}^T\widehat y_{k}\notag\\
	\leq &\ u_k^T\left(y_{k+1}-y_{k}\right)+\widehat u_k^T\left(\widehat y_{k+1}-\widehat y_{k}\right)-\epsilon\|\widehat y_{k+1}-\widehat y_{k}\|^2\notag\\
	&-y_{k+1}^T\widehat y_{k+1}+y_{k}^T\widehat y_{k}\notag\\
	=&\ \widehat y_{k+1}^T\left(y_{k+1}-y_{k}\right)+y_k^T\left(\widehat y_{k+1}-\widehat y_{k}\right)-\epsilon\|\widehat y_{k+1}-\widehat y_{k}\|^2\notag\\
	&-y_{k+1}^T\widehat y_{k+1}+y_{k}^T\widehat y_{k}\notag\\
	=& -\epsilon\|\widehat y_{k+1}-\widehat y_{k}\|^2\notag\\
	\leq &\ 0.
\end{align*}
The closed-loop system is stable in the sense of Lyapunov. Moreover, $W(x_{k},\widehat x_{k})-W(x_{k},\widehat x_{k}) = 0$ implies that $\widehat y_{k+1}-\widehat y_{k}=0$. We use LaSalle's invariance principle in the following; e.g., see Theorem 1 of Reference \cite{mei2017lasalle}. In the case that $W(x_{k},\widehat x_{k})-W(x_{k},\widehat x_{k})$ stays at zero for all future time steps $k$, we have $\widehat y_{k+1}-\widehat y_{k}\equiv 0$. Since $\widehat y_k$ is the output of the OSNI system associated with the SAOSNI system $\widetilde H_2$; i.e., $\widetilde y_k=\widehat y_{k+1}$, we also have that $\widetilde y_{k} = \widetilde y_{k-1}$ for all future time steps $k$. That is $\widetilde y_{k+1} = \widetilde y_{k}$ for all future time steps. In this case, according to the system setting $u_k=\widetilde y_k$, the input $u_k$ also remains constant and it equals to $\widetilde y_k$. According to Assumption \ref{assumption1}, this implies that $x_k$ and $\widehat x_{k}$ equal to zero after some time step $k^*$. Otherwise, the function  $W(x_k,\widehat x_k)$ will keep decreasing until $W(x_k,\widehat x_k)=0$. That is, $x_k=0$ and $\widehat x_k=0$, which is the equilibrium of the closed-loop interconnection. \hfill $\blacksquare$
\end{pf}

In a similar manner to the above, an OSNI system can be stabilized using an auxiliary system for an NI system, which we call a step advanced negative imaginary (SANI) system. We provide the definition of SANI systems as follows.
Consider the system 
\begin{subequations}\label{eq:bar H1}
\begin{align}
\overline H_1\colon \quad	x_{k+1} =&\ f(x_k,u_k),\\
	\overline y_k=&\ \overline h(x_k,u_k),
\end{align}	
\end{subequations}
where $f\colon\mathbb R^n \times \mathbb R^p\to \mathbb R^n$ and $\overline h\colon\mathbb R^n\times \mb R^p \to \mathbb R^p$. Here, $u_k, \overline y_k \in \mathbb R^p$ and $x_k\in \mathbb R^n$ are the input, output and state of the system, respectively, at time step $k\in \mb N$.

\begin{definition}\label{def:DT-SANI}
	The system \eqref{eq:bar H1} is said to be a discrete-time step advanced negative imaginary (SANI) system if there exists a continuous function $h(x_k)$ such that:
\begin{enumerate}
	\item $\overline h(x_k,u_k)=h(f(x_k,u_k))$;
	\item There exists a continuous positive definite storage function $V\colon \mb R^n \to \mb R$ such that for arbitrary states $x_k$ and inputs $u_k$,
\begin{equation*}
V(x_{k+1})-V(x_{k})\leq u_k^T\left(h(x_{k+1})-h(x_{k})\right)
\end{equation*}
for all $k$.
\end{enumerate}	
\end{definition}

\begin{figure}[h!]
\centering
\psfrag{in_0}{$r=0$}
\psfrag{in_1}{$u_k$}
\psfrag{y_1}{$\overline y_k$}
\psfrag{e}{$\widehat u_k$}
\psfrag{x_h}{$\widehat y_k$}
\psfrag{plant}{$\overline H_1$}
\psfrag{HIGS}{\hspace{0.2cm}$H_2$}
\includegraphics[width=8.5cm]{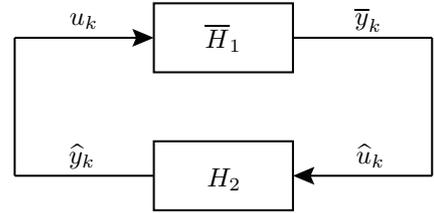}
\caption{Closed-loop interconnection of the SANI controller $\overline H_1$ and the OSNI plant $H_2$.}
\label{fig:interconnection2}
\end{figure}

An SANI system can be used in the control of an OSNI plant when they are interconnected as shown in Fig.~\ref{fig:interconnection2}.
The interconnection settings in Fig.~\ref{fig:interconnection2} are described by the following equations:
\begin{align*}
\widehat u_k =&\  \overline y_k;\\
u_k =&\ \widehat y_k.
\end{align*}

\begin{figure}[h!]
\centering
\psfrag{in_1}{$u_k$}
\psfrag{out_1}{$\overline y_k$}
\psfrag{in_2}{\hspace{0.1cm}$\widehat u_k$}
\psfrag{out_2}{\hspace{0.08cm}$\widehat y_k$}
\psfrag{sys1}{\hspace{0cm}$\overline H_1$}
\psfrag{sys2}{\hspace{0.03cm}$H_2$}
\includegraphics[width=8.5cm]{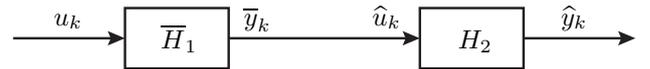}
\caption{Open-loop interconnection of the SANI plant $\overline H_1$ and the OSNI controller $H_2$.}
\label{fig:open-loop2}
\end{figure}

We make the following assumptions for the open-loop interconnection of the systems $\overline H_1$ and $H_2$ as shown in Fig.~\ref{fig:open-loop2}.

\begin{assumption}\label{assumption2}
	Giving an input $u_k=\overline u$ for the system $\overline H_1$ which is constant for all time steps $k\geq k_c$, we obtain a corresponding output $\overline y_k$. Also, setting the input of the system $H_2$ to be $\widehat u_k=\overline y_k$ in the open-loop interconnection shown in Fig.~\ref{fig:open-loop2} with corresponding output $\widehat y_k$. If $\widehat y_k=\overline {\widehat y}$ remains constant for all $k\geq k_c$, and $\overline {\widehat y}=\overline u$, then
$x_k,\widehat x_k=0$
for all $k\geq k^*$, where $k^*$ is some integer such that $k^*\geq k_c$.
\end{assumption}

\begin{theorem}\label{theorem:stability SANI and OSNI}
	Consider the SANI system $\overline H_1$ given by (\ref{eq:bar H1}) and the OSNI system $H_2$ given by (\ref{eq:H2}). Also, suppose Assumption \ref{assumption2} is satisfied and the function
\begin{equation*}\label{eq:W in stability SANI and OSNI}
W(x_k,\widehat x_k) = V(x_k)+\widehat V(\widehat x_k)-h(x_k)^T\widehat h(\widehat x_k)
\end{equation*}
is continuous and positive definite, where $V(x_k)$ and $\widehat V(\widehat x_k)$ are the storage functions of the systems $\overline H_1$ and $H_2$, respectively. Then the closed-loop interconnection of the system $\overline H_1$ and the system $H_2$ shown in Fig.~\ref{fig:interconnection2} is asymptotically stable.
\end{theorem}
\begin{pf}
	The proof of Theorem \ref{theorem:stability SANI and OSNI} follows a similar derivation to the proof of Theorem \ref{theorem:stability NI and SAOSNI}. \hfill $\blacksquare$
\end{pf}

\section{THE NI PROPERTY FOR LINEAR SYSTEMS}\label{sec:nonlinear DT-NI for linear}
In this section, we provide conditions under which a linear system is discrete-time NI according to Definition \ref{def:DT_NNI}. We also specialize the results in Section \ref{sec:stability} to linear systems. Consider the case where the system (\ref{eq:H1}) is linear and has a state-space realization as follows
\begin{subequations}\label{eq:DT linear system}
	\begin{align}
		x_{k+1}=&\ Ax_k+Bu_k,\\
		y_k =&\ Cx_k.
	\end{align}
\end{subequations}
Here $u_k,y_k \in \mathbb R^p$ and $x_k\in \mathbb R^n$ still denote the input, output and state of the system, respectively, at time step $k\in \mathbb N$. Also, $A\in \mb R^{n\times n}$, $B\in \mb R^{n\times p}$ and $C\in \mb R^{p\times n}$.

The following theorem provides necessary and sufficient LMI conditions under which the system (\ref{eq:DT linear system}) is NI, with a quadratic storage function.

\begin{theorem}\label{thm:LMI new DT-NI}
The linear system (\ref{eq:DT linear system}) is NI with a quadratic storage function of the form
\begin{equation}\label{eq:quadratic storage}
	V(x_k)= \frac{1}{2}x_k^T Px_k
\end{equation}
satisfying the dissipation inequality (\ref{eq:NNI ineq}), if and only if the following matrix is positive semidefinite:
\begin{equation}
\label{M}
M=\begin{bmatrix}
	P-A^TPA & (A^T-I)C^T -A^TPB\\ C(A-I) -B^TPA & CB+B^TC^T-B^TPB
\end{bmatrix}. 
\end{equation}
Furthermore, in the case when 
\begin{equation}\label{det}
    \det (I - A)\neq 0, 
\end{equation}
the NI property (\ref{eq:NNI ineq}) with the storage function (\ref{eq:quadratic storage}) for the system (\ref{eq:DT linear system}) is equivalent to the fulfillment of the following two conditions:
\begin{align}
\centering
\label{C}
  C & = 
  B^T (I-A)^{-T}P, 
  \\
\label{PP}
   & P - A^T P A  \geq 0. 
\end{align}
\end{theorem}
\begin{pf}
	In view of the system dynamics (\ref{eq:DT linear system}), the NI property (\ref{eq:NNI ineq}) can be represented by the inequality 
\begin{equation}
\label{diff0}
     u^T C(Ax+Bu - x)  - (V(Ax+Bu)-V(x)) \geq  0
\end{equation}
for any state $x \in \mb R^n$ and input $u \in \mb R^p$. This inequality is obtained by subtracting the left-hand side of (\ref{eq:NNI ineq}) from its right-hand side. In application to the quadratic storage function (\ref{eq:quadratic storage}), the condition  (\ref{diff0}) takes the form
\begin{align}
\nonumber
0  \leq &\ 
    u^T C((A-I)x+Bu)+ \frac{1}{2}x^TPx\notag\\
    &   -\frac{1}{2}(Ax+Bu)^TP(Ax+Bu)\notag \\
\nonumber
    = &\  
    \frac{1}{2}
    \left[x^T \left(P-A^T P A\right)\right.\\
\nonumber
    & \left.+ u^T \left(CB + B^T C^T - B^T P B\right)u\right]\\
\nonumber
    & + 
    u^T \left(C(A-I) - B^T PA\right)x \\
\label{diff1}
    = &\
    \frac{1}{2}
    \xi ^T M \xi, 
\end{align}
where $M$ is defined as in (\ref{M}) and the right-hand side is a quadratic function of the augmented state-input vector
\begin{equation}
\label{z}
    \xi \colonequals
    \begin{bmatrix}
      x \\
      u
    \end{bmatrix} , 
\end{equation}
which can take any value in $\mb R^{n+p}$.  
The inequality (\ref{diff1}) holds  for arbitrary $\xi$ if and only if the matrix $M$ is positive semi-definite:
\begin{equation}
\label{Mpos}
  M \geq 0. 
\end{equation}
Therefore, the NI property (\ref{eq:NNI ineq}) with the quadratic storage function (\ref{eq:quadratic storage}) is indeed equivalent to (\ref{Mpos}).  We will now prove that, under the condition (\ref{det}), the inequality (\ref{Mpos}) is equivalent to (\ref{C}) and (\ref{PP}). Note that (\ref{Mpos}) implies (\ref{PP}) regardless of (\ref{det}) because $P-A^T PA$ is a diagonal block of $M$. Now since (\ref{Mpos}) implies the NI property (\ref{diff0}), the latter holds in particular,  when the system state $x$ is related to an arbitrary input $u \in \mb R^p$ by
\begin{equation}
\label{xu}
    x = (I-A)^{-1}B u,
\end{equation}
where, this time,  we have used (\ref{det}). Since the relation (\ref{xu}) is equivalent to 
$    Ax+Bu = x $, 
any such pair $(x,u)$ makes the left-hand side of (\ref{diff0}) and the right-hand side of (\ref{diff1}) vanish. Therefore, the corresponding vector $\xi$ in (\ref{z}) takes the form 
\begin{align}
\label{zF}
    \xi =& 
    \begin{bmatrix}
      (I-A)^{-1}B u\\
      u 
    \end{bmatrix}
    =
    F u,\notag\\
\textnormal{where} \qquad    F \colonequals &
        \begin{bmatrix}
      (I-A)^{-1}B\\
      I
    \end{bmatrix}
    \in \mb R^{(n+p)\times p}, 
\end{align}
which satisfies $\xi^TM\xi = 0$ (recall that (\ref{Mpos}) holds), and hence, belongs to the null space of the positive semi-definite matrix $M$:
\begin{equation}
\label{MFu}
    0 = M\xi = MFu. 
\end{equation}
Since $u \in \mb R^p$ was arbitrary, the relation (\ref{MFu}) implies
\begin{equation}
\label{MF0}
    MF = 0.
\end{equation}
In particular, (\ref{MF0}) implies the product of the first block-row $M_{1\bullet}$ of the matrix $M$ in (\ref{M}) and the matrix $F$ from (\ref{zF}) vanish:
\begin{align*}
\nonumber
    0
    =&\
    M_{1\bullet} F\\
    =& \begin{bmatrix}
       P - A^T PA &(A^T-I)C^T -A^T P B
     \end{bmatrix}
     \begin{bmatrix}
      (I-A)^{-1}B\\
\nonumber
      I
    \end{bmatrix}     \\
\nonumber
    =&\
    (P - A^T PA) (I-A)^{-1}B
    +
    (A^T-I)C^T -A^T P B\\
\nonumber
    =&\ 
    P(I-A)^{-1}B - A^T P((I-A)^{-1}-I)B\\
    &+
    (A^T-I)C^T -A^T P B \notag\\        
\nonumber
    =&\
    P(I-A)^{-1}B - A^T P(I-A)^{-1}B
    +
    (A^T-I)C^T\\    
    =&\
    (A^T-I)(C^T - P(I-A)^{-1}B), 
\end{align*}
which leads to (\ref{C}) due to the nonsingularity of $A^T - I$ which follows from (\ref{det}). This completes the proof that (\ref{det}), (\ref{Mpos}) imply (\ref{C}), (\ref{PP}). It now remains to prove that (\ref{det}), (\ref{C}), (\ref{PP})  imply (\ref{Mpos}). To this end, we note that (\ref{det}) ensures (\ref{C}) is well-defined, and the substitution of (\ref{C}) into the appropriate blocks of (\ref{M}) yields
\begin{align}
\nonumber
    M_{22}
    =&\ 
    CB + B^T C^T - B^T P B\\
\nonumber
    =&\   
    B^T (I-A)^{-T}P B
    +
    B^T P (I-A)^{-1} B
    - B^T P B\\
\nonumber
    =&\
    B^T (I-A)^{-T}
    (
    P(I-A) 
    +
    (I-A^T)P \notag\\
    &
    -
    (I-A^T) P (I-A)) 
    (I-A)^{-1} B \notag\\
\label{M22}
    =&\
    B^T (I-A)^{-T}
    ( P - A^T P A)
    (I-A)^{-1} B,\\    
\nonumber
M_{12}
    =&\ 
(A^T-I)C^T  -A^T P B\\
\nonumber
    =&\     
    (A^T-I)P(I-A)^{-1}B  -A^T P B\\    
\nonumber
    =&\ 
    \left((A^T-I)P  -A^T P (I-A) \right)(I-A)^{-1}B\\    
\label{M12}
    =& 
    -(P-A^T P A)(I-A)^{-1}B.
\end{align}
By substituting (\ref{M22}) and (\ref{M12}) along with $M_{21} = M_{12}^T$ into (\ref{M}), it follows that the matrix $M$ can be factorized as
\begin{equation}
\label{MG}
  M
  =
  G^T
  (P-A^T P A)
  G,
  \end{equation}
where $G\colonequals 
  \begin{bmatrix}
    - I & (I-A)^{-1}B 
  \end{bmatrix}
  \in \mb R^{n\times (n+p)}$.
The factorization (\ref{MG}) shows that  (\ref{PP}) implies (\ref{Mpos}). Thus (\ref{det}), (\ref{C}), (\ref{PP})  imply (\ref{Mpos}). This completes the proof. \hfill $\blacksquare$
\end{pf}

We now provide frequency domain conditions for a linear system to be NI. The following preliminary result is required.
\begin{lemma}\cite{ferrante2017discrete}\label{lemma:original DT-NI conditions}
	Suppose the system (\ref{eq:DT linear system}) is minimal and satisfies $\det(I+A)\neq 0$ and $\det(I-A)\neq 0$. Then there exists $X=X^T>0$ satisfying
	\begin{equation*}\label{eq:LMI conditions for original DT-NI}
	X-A^TXA\geq 0,\ \textnormal{and}\ C = B^T(I-A)^{-T}X(A+I)	
	\end{equation*}
if and only if the transfer matrix $\wc G(z)=C(zI-A)^{-1}B$ satisfies the following conditions:
\begin{enumerate}
	\item $\wc G(z)$ has no poles in $|z|>1$;
	\item $i\left[\wc G(e^{i\theta})-\wc G(e^{i\theta})^*\right]\geq 0$ for all $\theta\in (0,\pi)$ except for the values of $\theta$ for which $z = e^{i\theta}$ is a pole of $\wc G(z)$;
	\item If $z_0 = e^{i\theta_0}$, with $\theta_0\in (0,\pi)$, is a pole of $\wc G(z)$, then it is a simple pole and the normalized residue matrix
	\begin{equation*}
		K_0\colonequals\frac{1}{z_0}\lim_{z\to z_0}(z-z_0)i\wc G(z)
	\end{equation*}
is Hermitian and positive semidefinite.
\end{enumerate}
\end{lemma}
\begin{pf}
	See Lemma 3.2 and Theorem 3.2 of \cite{ferrante2017discrete}. Note that the cases in which $G(z)$ has poles at $z_0=1$ and $z_0=-1$ are excluded by the assumptions $\det(I+A)\neq 0$ and $\det(I-A)\neq 0$.\hfill $\blacksquare$
\end{pf}
The following theorem provides frequency-domain conditions for a linear system to be NI.
\begin{theorem}\label{thm:frequency-domain conditions for novel DT-NI}
	Suppose the system (\ref{eq:DT linear system}) is minimal and satisfies $\det(I+A)\neq 0$ and $\det(I-A)\neq 0$. Then the system (\ref{eq:DT linear system}) is NI according to Definition \ref{def:DT_NNI} if and only if its transfer matrix $G(z)=C(zI-A)^{-1}B$ satisfies the following conditions:
	\begin{enumerate}
		\item $G(z)$ has no poles in $|z|>1$;
		\item $i\left[(e^{i\theta}+1)G(e^{i\theta})-(e^{-i\theta}+1)G(e^{i\theta})^*-L+L^T\right]\geq 0$ for all $\theta\in (0,\pi)$ except for the values of $\theta$ for which $z = e^{i\theta}$ is a pole of $G(z)$. Here,
	\begin{equation}\label{eq:L}
		L = \lim_{z\to \infty}zG(z);
	\end{equation}
	\item If $z_0 = e^{i\theta_0}$, with $\theta_0\in (0,\pi)$, is a pole of $G(z)$, then it is a simple pole and the normalized residue matrix
	\begin{equation}\label{eq:new K_0}
		K_0\colonequals \left(1+\frac{1}{z_0}\right)\lim_{z\to z_0}(z-z_0)iG(z)
	\end{equation}
	is Hermitian and positive semidefinite.
	\end{enumerate}
\end{theorem}
\begin{pf}
	By comparing the LMI conditions in Theorem \ref{thm:LMI new DT-NI} and Lemma \ref{lemma:original DT-NI conditions}, we have that the system (\ref{eq:DT linear system}) is NI if and only if the transfer matrix $\wc  G(z)$ of the realization $(A,B,C(A+I))$ satisfies the frequency-domain conditions (1)--(3) in Lemma \ref{lemma:original DT-NI conditions}. By letting $\wc y_k = C(A+I)x_k$, we have
	\begin{equation*}
		\wc y_k = CAx_k+Cx_k = y_{k+1}+y_k-CBu_k.
	\end{equation*}
This implies that
	\begin{equation}\label{eq:two G(z) relationship}
		\wc G(z) = (z+1)G(z)-CB.
	\end{equation}
The matrix $CB$ can be represented as $CB = \lim_{z\to \infty}zG(z)$. We show in the following that $\wc G(z)$ satisfies Conditions (1)--(3) in Lemma \ref{lemma:original DT-NI conditions} if and only if $G(z)$ satisfies Conditions (1)--(3) in the present Theorem. We substitute (\ref{eq:two G(z) relationship}) into the corresponding conditions in Lemma \ref{lemma:original DT-NI conditions}. More precisely, the transfer matrix $\wc G(z)$ satisfies Condition (1) in Lemma \ref{lemma:original DT-NI conditions} if and only if Condition (1) in the present theorem is satisfied.
Condition (2) is established by substituting (\ref{eq:two G(z) relationship}) into $i\left[\wc G(e^{i\theta})-\wc G(e^{i\theta})^*\right]\geq 0$.
Condition (3) is established via the following derivation
\begin{align*}
	K_0 =& \frac{1}{z_0}\lim_{z\to z_0}(z-z_0)i\wc G(z)\notag\\
	=& \frac{1}{z_0}\lim_{z\to z_0}(z-z_0)i\left[(z+1)G(z)-CB\right]\notag\\
	=& \frac{z_0+1}{z_0}\lim_{z\to z_0}(z-z_0)iG(z).
\end{align*}
\hfill $\blacksquare$
\end{pf}

We provide the following two corollaries of Theorems \ref{theorem:stability NI and SAOSNI} and \ref{theorem:stability SANI and OSNI}, which specialize the stability results to linear systems. We show in the following that the positive definiteness conditions in Theorems \ref{theorem:stability NI and SAOSNI} and \ref{theorem:stability SANI and OSNI} are reduced to DC gain conditions.

Consider a minimal system
	\begin{equation}\label{eq:Corollary ABC}
		x_{k+1} = A_1x_k+B_1u_k; \quad y_k = C_1x_k
	\end{equation}
where $x_k\in \mathbb R^n$, $u_k,y_k\in \mathbb R^p$ is the state, input and output, respectively. Here, 
$A_1\in \mb R^{n\times n}$, $B_1\in \mb R^{n\times p}$ and $C_1\in \mb R^{p\times n}$.

Consider another minimal system
	\begin{equation}\label{eq:Corollary1 ABCD}
		\widetilde x_{k+1} = A_2\widetilde x_k+ B_2 \widetilde u_k; \quad \widetilde y_k = C_2 \widetilde x_k + D_2 \widetilde u_k
	\end{equation}
where $\widetilde x_k\in \mathbb R^m$, $\widetilde u_k,\widetilde y_k\in \mathbb R^p$ is the state, input and output, respectively. Here, 
$A_2\in \mb R^{m\times m}$, $B_2\in \mb R^{m\times p}$ and $C_2\in \mb R^{p\times m}$.

\begin{corollary}\label{corollary1}
Consider a minimal NI system of the form (\ref{eq:Corollary ABC}) with transfer matrix $G(z)$ and a minimal SAOSNI system of the form (\ref{eq:Corollary1 ABCD})
 with transfer matrix $H(z)$. Suppose $\lambda_{max}(G(1)H(1))<1$, then the closed-loop interconnection of the systems (\ref{eq:Corollary ABC}) and (\ref{eq:Corollary1 ABCD}) is asymptotically stable. 
\end{corollary}
\begin{pf}
Since the system (\ref{eq:Corollary1 ABCD}) is SAOSNI, then according to Definition \ref{def:DT-SAOSNI}
there exists an OSNI system of the form
\begin{equation}\label{eq:Corollary OSNI}
	\widetilde x_{k+1} = A_2\widetilde x_k+ B_2 \widetilde u_k; \quad \widehat y_k = \widehat C_2 \widetilde x_k
\end{equation}
where $C_2 = \widetilde C_2A_2$ and $D_2 = \widetilde C_2B_2$. In other words, $\widetilde y_k = \widehat y_{k+1}$. Due to the step advance, the system (\ref{eq:Corollary OSNI}) has transfer matrix $\widehat H(z)$ such that $H(z) = z\widehat H(z)$. Therefore, $\widehat H(1) = H(1)$. According to Theorem \ref{theorem:stability NI and SAOSNI}, a Lyapunov function for the closed-loop interconnection of the systems (\ref{eq:Corollary ABC}) and (\ref{eq:Corollary1 ABCD}) is given by
	\begin{equation}\label{eq:Corollary W}
		W(x,\widetilde x) = \begin{bmatrix}
			x^T & \widetilde x^T
		\end{bmatrix}\begin{bmatrix}
			P & -C_1^T\widehat C_2 \\ -\widehat C_2^TC_1 & \widetilde P
		\end{bmatrix}\begin{bmatrix}
			x \\ \widetilde x
		\end{bmatrix}
	\end{equation}
where $x$ and $\widetilde x$ are the states of the state-space realizations of $G(z)$ and $\widehat H(z)$, respectively. Also, $P$ and $\widetilde P$ are symmetric positive definite matrices satisfying condition (\ref{C}) for the state-space realizations of $G(z)$ and $\widehat H(z)$, respectively. Note that the existence of such $P$ and $\widetilde P$ follows from the NI properties of $G(z)$ and $H(z)$ according to Theorem \ref{thm:LMI new DT-NI}. We prove that if $\lambda_{max}(G(1)H(1))<1$, then the Lyapunov function given by (\ref{eq:Corollary W}) is positive definite. We have that
\begin{align*}
	G(1)H(1) =&\ G(1)\widehat H(1)\notag\\
	 =&\ C_1(I-A_1)^{-1}B_1\widehat C_2(I-A_2)^{-1}B_2\notag\\
	=&\ C_1P^{-1}C_1^T\widehat C_2\widetilde P^{-1}\widehat C_2^T,
\end{align*}
where the last equality follows from condition (\ref{C}). Therefore, $\lambda_{max}(G(1) H(1))<1$ implies
\begin{equation*}
	\lambda_{max}(P^{-1}C_1^T\widehat C_2\widetilde P^{-1}\widehat C_2^TC_1)<1.
\end{equation*}
This implies that 
\begin{equation}\label{eq:P>C1C2P-1C2C1}
	P>C_1^T\widehat C_2\widetilde P^{-1}\widehat C_2^TC_1.
\end{equation}
By considering the Schur's complement, inequality (\ref{eq:P>C1C2P-1C2C1}) and the fact $\widetilde P>0$ imply that $\begin{bmatrix}
			P & -C_1^T\widehat C_2 \\ -\widehat C_2^TC_1 & \widetilde P
		\end{bmatrix}>0$. Hence, $W(x,\widetilde x)$ given in (\ref{eq:Corollary W}) is positive definite. Note that in this case, Assumption \ref{assumption1} is satisfied because of the minimality of the systems (\ref{eq:Corollary ABC}) and (\ref{eq:Corollary1 ABCD}) and also the assumption $\lambda_{max}(G(1)H(1))<1$. The rest of the proof follows from Theorem \ref{theorem:stability NI and SAOSNI}.\hfill $\blacksquare$
\end{pf}

\begin{corollary}
Consider a minimal OSNI system of the form (\ref{eq:Corollary ABC}) with transfer matrix $G(z)$ and a minimal SANI system of the form (\ref{eq:Corollary1 ABCD}) with transfer matrix $H(z)$. Suppose $\lambda_{max}(G(1)H(1))<1$, then the closed-loop interconnection of the systems (\ref{eq:Corollary ABC}) and (\ref{eq:Corollary1 ABCD}) is asymptotically stable. 
\end{corollary}
\begin{pf}
	The proof is similar to that of Corollary \ref{corollary1}, with Theorem \ref{theorem:stability SANI and OSNI} used instead of Theorem \ref{theorem:stability NI and SAOSNI}.\hfill $\blacksquare$
\end{pf}

\begin{remark}\label{remark:SANI frequency conditions}
	For the case of a linear SANI system with transfer matrix $H(z)$, the conditions in Definition \ref{def:DT-SANI} reduce to the condition that $H(z)=zH_1(z)$ where $H_1(z)$ is an NI system. In other words, suppose $H(z)$ has no pole at $1$ or $-1$, then it is SANI if and only if $(1+\frac{1}{z})H(z)-\lim_{z\to \infty}H(z)$ satisfies Conditions (1)--(3) in Lemma \ref{lemma:original DT-NI conditions}. 
\end{remark}

\section{CONCLUSION}\label{sec:conclusion}
This paper provides a new definition of the NI property for discrete-time systems. According to this definition, ZOH sampling of a continuous-time NI system yields a discrete-time NI system. We prove that the closed-loop interconnection of an NI system and an SAOSNI system is asymptotically stable, under certain assumptions. Under similar assumptions, asymptotic stability is also proved for the closed-loop interconnection of an SANI system and an OSNI system. We provide necessary and sufficient LMI conditions and frequency-domain conditions for a linear system to be NI. The stability results proposed for general nonlinear systems are also specialized for linear systems.



\end{document}